\documentclass[useAMS,usenatbib]{mn2e}
\usepackage{graphicx}

\newcommand{\rerevised}[1]{{ #1}}
\newcommand{\revised}[1]{{ #1}}

\title[HFQPO in black-hole binaries]{High-Frequency Quasi-Periodic Oscillations in black-hole binaries}
\author[T. M. Belloni, A. Sanna and M. M\' endez]{T.M. Belloni$^{1}$\thanks{E-mail:
tomaso.belloni@brera.inaf.it}, A. Sanna$^{2}$, M. M\'endez$^{2}$\\
$^{1}$INAF - Osservatorio Astronomico di Brera, Via E. Bianchi 46, I-23807, Merate, Italy\\
$^{2}$Kapteyn Astronomical Institute, University of Groningen, P.O. Box 800, 9700 AV Groningen, The Netherlands}
\begin{document}

\date{Accepted 2012 Month day. Received 2012 Month day; in original form 2012 Month day}

\pagerange{\pageref{firstpage}--\pageref{lastpage}} \pubyear{2012}

\maketitle

\label{firstpage}

\begin{abstract}
We present the results of the analysis of a large database of X-ray observations of 22 galactic black-hole transients with the Rossi X-Ray timing explorer throughout its operative life for a total exposure time of $\sim$12 Ms. We excluded persistent systems and the peculiar source GRS 1915+105, as well as the most recently discovered sources. The semi-automatic homogeneous analysis was aimed at the detection of high-frequency (100-1000 Hz) quasi-periodic oscillations (QPO), of which \revised{several} cases were previously reported in the literature.
After taking into account the number of independent trials, we obtained 11 detections from two sources only: XTE J1550-564 and GRO J1655-40. For the former, the detected frequencies are clustered around 180 Hz and 280 Hz, as previously found. For the latter, the previously-reported dichotomy  300-450 Hz is found to be less sharp.
We discuss our results in comparison with kHz QPO in neutron-star X-ray binaries and the prospects for future timing X-ray missions.

\end{abstract}

\begin{keywords}
accretion, accretion discs -- black hole physics -- relativistic processes -- X-rays: binaries
\end{keywords}

\section{Introduction}

The Rossi X-Ray Timing Explorer (RXTE) mission has provided thousands of high-quality observations of black-hole transients (BHTs), which have profoundly changed our knowledge of the properties of accretion onto stellar-mass black holes (see e.g. Belloni 2010; Fender 2010; Belloni et al. 2012). An important aspect has been the detection of Quasi-Periodic Oscillations (QPO) at frequencies higher than $\sim$40 Hz, which opened a new window onto fast phenomena in the frequency range expected from signals associated to Keplerian motion in the innermost regions of an accretion disk around a black hole. A small number of detections is available in the literature from a \revised{a small number of} sources: GRS 1915+105 (Morgan et al. 1997; Strohmayer 2001a; Belloni et al. 2001; Remillard et al. 2002b; Belloni et al. 2006), GRO J1655-40 (Remillard et al. 1999; Strohmayer 2001b), XTE J1550-564 (Homan et al. 2001; Miller et al. 2001; Remillard et al. 2002a), H1743-322 (Homan et al. 2005; Remillard et al. 2006), XTE J1650-500 (Homan et al. 2003), 4U 1630-47 (Klein-Wolt, Homan \& van der Klis 2004), XTE J1859+226 (Cui et al. 2000) and IGR J17091-3624 (Altamirano \& Belloni 2012). The signals are very weak and some detections appear to have marginal statistical significance. Although these High-Frequency QPOs (hereafter HFQPOs) are very important as they constitute a promising way to detecting effects of General Relativity in the strong field regime, no systematic work on the existing bulk of RXTE data is available. In particular, it is important to compare quantitatively the existing results with those from kilohertz QPOs from neutron star X-ray binaries, where signals are stronger and the known phenomenology much better covered (see e.g. van der Klis 2006 for a review). The similarities between the timing features of these two classes of sources are strong (see Wijnands et al. 1999; Psaltis et al. 1999; Belloni et al. 2002; van der Klis 2006), but it is not clear how HFQPOs fit in this comparison.
In this paper, we present a systematic analysis of a large sample of RXTE observations of BHTs, concentrating on the detection of HFQPOs. All major sources are present, with the exception of GRS 1915+105, which will be the subject of a separate paper due to the large number of available observations and its peculiarities) and IGR J17091-3624, which was observed by RXTE after the analysis part of this work was completed. An a posteriori bulk analysis of the data suffers from problems related to number of trials, which means that the sensitivity to detections will be reduced. In addition, our procedure was aimed only at detecting one peak in each Power Density Spectrum (PDS), which means that double detections do not appear. Finally, different observations were analyzed separately without any merging, although merging was the way some of these oscillations were found. However, such a (semi-automatic) procedure is useful to assess what is available from RXTE data and to plan future observations with upcoming facilities like ASTROSAT and LOFT. 
\revised{The comparison with previous works is in many cases difficult as the starting conditions for the analysis are markedly different. Motivated selection of observations and/or merging, as well as the relaxation of the constraints imposed in our search (total number of available observations, frequency range considered, energy range for the accumulation, limits on the quality factor of the signal) can lead to more sensitive results.
}

\begin{table*}
 \centering
 \begin{minipage}{170mm}
  \caption{Statistics of available data and obtained detections. Columns are: source name, total number of observations, number of observations analyzed in each band, total exposure considered in each band, number of observations with single try probability $>$3$\sigma$ in each band, number of detections after manual analysis in each band and final number of detections.}
  \begin{tabular}{@{}lcccccccccc@{}}
  \hline
   Source name  &
   N. obs.            &
   Valid T.            &
   Valid H.            &
   Exp. T (ks)         &
   Exp. H (ks)         &
   $>3\sigma$ T     &
   $>3\sigma$ H     &
   Good T              &
   Good H             &
   Final                  \\ 
 \hline
 GX 339-4               & 1353  & 893 & 517 & 2190 & 1657 & 31 & 34 &  4 & 11   &  4\\
 4U 1630-47            & 1002  & 846 & 701 & 1934 & 1672 & 31 & 27 & 11 & 7    &  4\\
 GRO J1655-40       &  596   & 516 & 491 & 2295 & 2214 & 18 & 24 &  7 &  8   & 11\\
 H 1743-322            &  504   & 386 & 300 & 1153 &  941  & 19 & 20 &  6 &  8   &  4\\
 XTE J1550-564      &  409   & 319 & 280 &  860 &  790   & 46 & 31 & 15 & 14 & 16\\
 Swift J1753.5-0127 & 278   & 240 & 172 &  671 &  527   & 14 &  8 &  3 &  3    &   0\\
 XTE J1752-223       & 208   & 138 &  79  &  343 &  250   &  6 &  6 &  0 &  4     &  2\\
 XTE J1650-500       & 182   & 101 &  69  & 219  &  155   & 17 &  5&  4 &  2    &   0\\
 Swift J1539.2-6227 & 156   &  61  &  42  &  153 &  122   &  1 &  3 &  1 &  1    &   0\\
 XTE J1817-330       & 155   & 112 &  79  &  372 &  256   &  3 &  2 &  1 &  1    &   0\\
 XTE J1859+226      & 131   & 115 & 102 &  314  & 281   &  6 &  4 &  2 &  2    &  1\\
 4U 1543-47             & 104   &   53 &  38  &  161  & 125   &  6 &  3 &  0 &  1    &   0\\
 XTE J1720-318       & 100   &   75 &  19  &  233  & 101   &  1 &  1 &  0  &  1    &   0\\
 4U 1957+115          & 100   &   97 &  46   & 529  & 334   &  1 &  0 &  0 &   0    &   0\\
 XTE J1118+480      &  94    &   62 &  50   &  174  & 154  &  0 &  3 &  0 &  1    &   0\\
 MAXI J1659-152     & 66     &   63 &  54   &  148  & 135  &  1 &  1 &  0 &  1    &   0\\
 SLX J1746-331       & 65     &   47 &  23   &  125  &   72  &  2 &  2 &  0 &  0    &   0\\
 XTE J1652-453       & 57     &   30 &    1   &    67  &    2   &  1 &  0 &  0 &  0    &   0\\
 XTE J1748-288       & 24     &   23 &  21   &  103  &   99  &  0 &  1 &  0 &  0    &   0\\
 SAX J1711.6-3808  & 17     &   17 &  12   &    43  &   30  &  0 &  0 &  0 &  0    &   0\\
 GS 1354-644          & 8       &     7  &    7   &   52  &    52  &  0 &  1 &  0 &  0    &   0\\
 GRS 1737-31          & 5       &     4  &   3   &   40  &    28  &  0 &  0 &  0 &  0    &   0\\
 \hline
 Total                        & 7108 &  4205 & 3106& 12177&9996 & 204& 176 & 54 & 65  & 42\\
 
\hline
\end{tabular}
\label{tab:obslog}
\end{minipage}
\end{table*}

\section[]{Data and analysis}

We selected all RXTE observations of known transient black-hole binaries available in the archive from the 
start of the mission until MJD 55601 (2011 February 9), concentrating on the data from the Proportional Counter Array (PCA) instrument. We analyzed 22 sources, for a total of 7108 observations (see Tab. \ref{tab:obslog}). The bright source GRS 1915+105 was not included for two reasons: its peculiarity and the fact that effectively during the RXTE lifetime it behaved as a persistent source. The analysis of data from GRS 1915+105 will be reported in a forthcoming paper.
For the project presented here, it is important to establish a detailed analysis procedure in advance, in order to control and 
minimize the probability of spurious detections. The following multi-step procedure was applied.

\begin{enumerate}

\item Two energy bands were considered for analysis: absolute channels 0-79 and 14-79 (corresponding to energies 1.51-21.30 keV and 4.07-21.30 keV at the beginning of the mission and to energies 2.06-33.43 keV and 6.12-33.43 keV a the end of the mission, due to gain changes in the detectors). We will refer to these as the total and hard band respectively. For some observations, these exact bands were not available due to the data modes, in which case we chose the closest approximation to those boundaries. In a few cases, the data modes were such that we used the full channel range and/or the total and hard band resulted to be identical.
\revised{The choice of the hard band was made in order not to deplete too much the number of counts and allow the examination of more observations (see next bullet).}

\item When analyzing a large number of observations, the issue of number of trials must be considered carefully. Of all observations in our sample, a sizable number were of a source at the very start or at the end of an outburst, where count rate is low and the sensitivity to detect quasi-periodic features is very low. We decided not to analyze observations where the data are not expected to yield a positive detection for HFQPO. Rather than selecting a simple threshold in count rate to exclude the observations, we used the expression to estimate the significance (in number of sigmas, $n_\sigma$) of a broad feature in frequency domain

$$n_\sigma = {1\over 2} r_s^2 {S^2\over (S+B)} \sqrt{T_{exp}\over \Delta\nu}$$

(see van der Klis 1998) where $r_s$ is the fractional rms variability of the feature, $S$ is the source net count rate, $B$ is the corresponding background count rate,  $T_{exp}$ is the exposure time and $\Delta\nu$ is the FWHM of the feature. We assumed a QPO with 6\% fractional rms and 10 Hz FWHM and used the observed count rates and exposure time to estimate the expected sensitivity. We then excluded all hard or total observations which yielded $n_\sigma < 3$.

With this procedure we retained 4205 total observations and 3106 hard observations, for a cumulative exposure time of 12.2 Ms and 10.0 Ms respectively (see Tab.  \ref{tab:obslog}). We consider a separate observation (or dataset) the data corresponding to a RXTE observation ID.

\item The large number of observations made it necessary to perform an automatic search in order to select candidates for HFQPOs. For each observation, the adopted procedure, applied both to the total and hard energy band, was the following:

\begin{itemize}

\item We produced a PDS by dividing the dataset in intervals of 16s duration and averaging the corresponding PDS. The time resolution for all spectra was 4096 points per second, corresponding to a Nyquist frequency of $\sim$2 kHz. The PDS were normalized according to Leahy et al. (1983). The powers were rebinned logarithmically in such a way that each frequency bin was larger than the previous one by $\sim$2\%.
\revised{Uncertainties in the power values were estimated following van der Klis (1988).}

\item Since the PDS of BHTs are rather complex and we were interested only in the high-frequency region, we considered only the frequency range 100-1000 Hz. This means that our implicit definition of HFQPO is limited to centroid frequencies larger than 100 Hz.
\item We fitted the PDS with an automatic procedure with a model consisting of a power law (to account for the Poissonian noise component) and a Lorentzian. The slope of the power law was not fixed to 0 (constant component) in order to fit any possible remaining tail of low-frequency features. We did the fit by first fixing the Lorentzian centroid frequency to all values between 100 and 1000 Hz in steps by 1 Hz. The FWHM of the Lorentzian was bound between 0.5 Hz and 1000 Hz. A fit with a free centroid was then made around the frequency corresponding to the minimum chi square. This resulted in a single best fit for each observation. Notice that this procedure is aimed at detecting a single HFQPO in each PDS and would not be able to detect multiple peaks.

\item From each fit, we estimated the significance of the detection by dividing the normalization (integral) of the Lorentzian by its \revised{1$\sigma$ error on the negative side.}
Only features with this figure of merit larger than 3 were considered. The surviving number of observations can be seen in the $3\sigma$ columns of Tab. \ref{tab:obslog}.

\item Examining the best fit parameters, we were able to eliminate obvious meaningless fits. Those included fits where the FWHM had reached one of the two imposed limits (see above), making the feature impossibly narrow or broad, and those where the centroid frequency was close to the search limits. In addition, we discarded all features with a quality factor $Q=\nu_0 / FWHM$ significantly less than 2, as we are interested only in peaked signals.
We checked all these cases visually in order not to miss peculiar features. None were found. In this way, the numbers of HFQPO candidates dropped to those in the ``good'' columns in Tab. \ref{tab:obslog}.

\item For all remaining PDS, we made a manual fit with the same model described above and we re-established the significance of the fit in the same way. In case we found a peak at a consistent frequency, both in the hard and total PDS, we retained the one with the highest significance. The final numbers of detections from this procedure are shown in the last column of Tab. \ref{tab:obslog}. The total number was 42. \revised{For all fits, the reduced chi square was close to unity.}

\end{itemize}

\item The resulting HFQPOs have a significance (estimated in the way described above) larger than 3$\sigma$ for a single trial. We searched them in a number of PDS for each source and we considered a broad range of frequencies. Both these procedures translate in a number of trials. For each source, we considered how many PDS were searched (for the total plus hard band) and how many independent frequencies were considered: for each detection we divided the 900 Hz interval used for the search by the FWHM of the detected QPO. Keeping these two effects in mind of course lowers the significance.

We did this under the assumption that any frequency within the considered range would be accepted. If one wants to limit to certain frequencies, for instance considering only peaks around 180 Hz and 280 Hz for XTE J1550-564 as previously detected (Homan et al. 2001; Miller et al. 2001; Remillard et al. 2002a), the number of trials will change and so will the significance of the detections. For this reason, in Tab.  \ref{tab:results} we show all 42 detections. In the following, we will restrict the discussion only to those with a high significance.We  marked in boldface in the table the HFQPOs with a final chance probability less than $10^{-2}$. Notice that a number of detections have a (rounded) final chance probability of 1, making them unlikely to be real. One has a rounded probability of 0, as it was so low that the spreadsheet we used could not approximate it.

\item Notice that our procedure can only detect one peak per PDS. In principle, two different features can be detected in the total and hard PDS, but this did not happen. No clear additional peaks were seen visually in the PDS containing a detection.

\end{enumerate}

In addition, we produced a Hardness-Intensity Diagram (HID) for each source, one point per observation, by accumulating net energy spectra from PCU2 (the best calibrated of the five units of the PCA) using standard procedures \footnote{http://heasarc.nasa.gov/docs/xte/data\_analysis.html}. The count rates were corrected for gain changes by dividing the spectra by a Crab spectrum (power law with photon index 2 and normalization 10 ph/cm$^{-2}$s$^{-1}$keV$^{-1}$ at 1 keV) simulated with the detector response at the day of the observation. Intensity is accumulated over PCA channels 8-49, corresponding to 3.2-19 keV at the end of the mission. Hardness is defined as $H/S$ where $S$ and $H$ are the corrected rates in the channel bands 8-14 (3.2-6 keV) and 15-24 (6-10 keV) respectively.

\begin{table*}
 \centering
 \begin{minipage}{140mm}
  \caption{List of detected HFQPOs. Columns are: observation date (MJD), exposure time, band in which the QPO was detected (Total/Hard), centroid frequency, FWHM and fractional rms of the QPO, single trial and final chance probabilities, source observed count rate (in PCU2) and X-ray hardness.}
  \begin{tabular}{@{}lccccccccc@{}}
  \hline
   MJD  &
   Exp. (s) &
   B&
   $\nu_0$ (Hz)&
   FWHM (Hz)&
   \%rms &
   P$_0$&
   P$_F$&
   Rate             &
   HR                  \\ 
 \hline
 \multicolumn{10}{c}{XTE J1550-564}\\
 \hline
51076.000	&	2960	&	T	&	{\bf 180.76}	$_{-	4.37	}^{+	3.92	}$&	95.12	$_{-	22.00	}^{+	16.12	}$&	1.15	$_{-	0.12	}^{+	0.14	}$&	2.87E-07	&	1.62E-03	&	15700	&	0.68	\\
51101.607	&	1584	&	T	&	140.70	$_{-	1.92	}^{+	1.71	}$&	16.51	$_{-	4.84	}^{+	19.85	}$&	1.12	$_{-	0.17	}^{+	0.59	}$&	6.64E-04	&	1.00E+00	&	3135	&	0.57	\\
51108.076	&	9872	&	T	&	{\bf 182.27}	$_{-	3.20	}^{+	3.25	}$&	77.88	$_{-	9.77	}^{+	11.65	}$&	1.74	$_{-	0.16	}^{+	0.20	}$&	1.90E-08	&	1.31E-04	&	3636	&	0.53	\\
51115.281	&	2048	&	H	&	272.43	$_{-	4.09	}^{+	3.42	}$&	30.59	$_{-	8.95	}^{+	16.30	}$&	4.02	$_{-	0.47	}^{+	0.78	}$&	7.80E-06	&	1.28E-01	&	1785	&	0.41	\\
51241.802	&	3104	&	H	&	{\bf 284.07}	$_{-	1.36	}^{+	1.35	}$&	26.86	$_{-	3.12	}^{+	3.88	}$&	3.66	$_{-	0.17	}^{+	0.21	}$&	8.77E-29	&	0.00E+00	&	4258	&	0.42	\\
51242.507	&	1504	&	H	&	{\bf 280.00}	$_{-	2.49	}^{+	2.87	}$&	30.00	$_{-	6.85	}^{+	9.80	}$&	3.78	$_{-	0.32	}^{+	0.46	}$&	2.05E-09	&	3.69E-05	&	4107	&	0.43	\\
51245.354	&	2752	&	T	&	180.75	$_{-	2.54	}^{+	2.02	}$&	16.76	$_{-	6.21	}^{+	8.71	}$&	1.08	$_{-	0.13	}^{+	0.25	}$&	8.54E-06	&	2.40E-01	&	4718	&	0.54	\\
51247.979	&	2816	&	T	&	185.20	$_{-	4.93	}^{+	6.73	}$&	46.04	$_{-	12.13	}^{+	15.69	}$&	1.32	$_{-	0.17	}^{+	0.30	}$&	3.30E-05	&	3.21E-01	&	4509	&	0.54	\\
51255.158	&	4080	&	H	&	{\bf 280.84}	$_{-	1.92	}^{+	2.00	}$&	24.37	$_{-	6.28	}^{+	8.53	}$&	4.35	$_{-	0.41	}^{+	0.50	}$&	7.20E-08	&	1.59E-03	&	2329	&	0.38	\\
51258.497	&	1104	&	H	&	276.02	$_{-	6.20	}^{+	6.64	}$&	36.77	$_{-	12.36	}^{+	23.00	}$&	5.22	$_{-	0.76	}^{+	0.01	}$&	2.91E-04	&	9.86E-01	&	1816	&	0.35	\\
51259.253	&	880	&	H	&	274.87	$_{-	3.61	}^{+	4.30	}$&	26.00	$_{-	8.68	}^{+	14.57	}$&	5.09	$_{-	0.67	}^{+	1.00	}$&	7.53E-05	&	7.90E-01	&	1710	&	0.39	\\
51291.184	&	1536	&	T	&	134.75	$_{-	1.45	}^{+	1.76	}$&	7.81	$_{-	3.25	}^{+	5.16	}$&	6.84	$_{-	1.09	}^{+	1.28	}$&	8.45E-04	&	1.00E+00	&	88	&	0.42	\\
51664.409	&	2240	&	H	&	{\bf 274.40}	$_{-	3.47	}^{+	3.43	}$&	41.96	$_{-	10.66	}^{+	19.88	}$&	6.66	$_{-	0.64	}^{+	1.23	}$&	1.11E-07	&	1.42E-03	&	1347	&	0.29	\\
51664.637	&	2864	&	H	&	{\bf 274.97}	$_{-	3.21	}^{+	2.90	}$&	33.51	$_{-	8.16	}^{+	13.97	}$&	5.42	$_{-	0.51	}^{+	0.87	}$&	4.65E-08	&	7.47E-04	&	1419	&	0.31	\\
51665.406	&	2144	&	H	&	264.66	$_{-	3.19	}^{+	7.09	}$&	42.87	$_{-	15.43	}^{+	22.01	}$&	6.50	$_{-	1.02	}^{+	1.35	}$&	6.87E-04	&	1.00E+00	&	1105	&	0.28	\\
51668.829	&	2976	&	H	&	263.29	$_{-	9.96	}^{+	9.41	}$&	53.90	$_{-	17.58	}^{+	30.55	}$&	5.50	$_{-	0.75	}^{+	1.56	}$&	1.36E-04	&	7.44E-01	&	856	&	0.26	\\
\hline
 \multicolumn{10}{c}{GRO J1655-40}\\
 \hline
50296.311	&	9056	&	T	&	{\bf 273.72}	$_{-	3.96	}^{+	3.97	}$&	70.33	$_{-	10.92	}^{+	19.40	}$&	1.15	$_{-	0.08	}^{+	0.15	}$&	1.15E-13	&	1.483E-9	&	7443	&	0.67	\\
50301.665	&	6464	&	T	&	307.16	$_{-	6.77	}^{+	8.17	}$&	68.46	$_{-	38.07	}^{+	0.00	}$&	0.88	$_{-	0.11	}^{+	0.21	}$&	3.91E-05	&	4.04E-01	&	6414	&	0.61	\\
50311.391	&	1856	&	T	&	285.56	$_{-	11.19	}^{+	8.91	}$&	63.78	$_{-	17.38	}^{+	42.48	}$&	1.14	$_{-	0.16	}^{+	0.36	}$&	1.93E-04	&	9.35E-01	&	5450	&	0.54	\\
50317.441	&	6176	&	H	&	443.16	$_{-	3.79	}^{+	4.12	}$&	41.44	$_{-	13.66	}^{+	20.27	}$&	5.39	$_{-	0.64	}^{+	0.81	}$&	1.28E-05	&	2.44E-01	&	5210	&	0.51	\\
50324.380	&	4896	&	T	&	{\bf 289.67}	$_{-	6.03	}^{+	6.26	}$&	98.78	$_{-	17.67	}^{+	40.67	}$&	1.41	$_{-	0.11	}^{+	0.30	}$&	1.59E-10	&	1.46E-06	&	7267	&	0.65	\\
50330.254	&	6144	&	H	&	453.72	$_{-	5.32	}^{+	6.29	}$&	37.69	$_{-	12.16	}^{+	16.46	}$&	5.69	$_{-	0.71	}^{+	0.89	}$&	2.91E-05	&	5.03E-01	&	4964	&	0.49	\\
50335.913	&	8160	&	H	&	{\bf 446.35}	$_{-	4.15	}^{+	4.10	}$&	38.74	$_{-	8.00	}^{+	9.80	}$&	6.03	$_{-	0.54	}^{+	0.59	}$&	9.01E-09	&	2.11E-04	&	4983	&	0.50	\\
50383.565	&	6192	&	H	&	442.34	$_{-	4.28	}^{+	4.71	}$&	31.50	$_{-	10.61	}^{+	15.62	}$&	4.55	$_{-	0.62	}^{+	0.70	}$&	1.31E-04	&	9.77E-01	&	6092	&	0.54	\\
50394.884	&	5792	&	T	&	{\bf 313.45}	$_{-	9.81	}^{+	9.38	}$&	98.11	$_{-	19.10	}^{+	38.42	}$&	1.17	$_{-	0.11	}^{+	0.25	}$&	3.72E-08	&	3.44E-04	&	5497	&	0.59	\\
53498.415	&	13968	&	T	&	125.81	$_{-	1.20	}^{+	1.10	}$&	7.49	$_{-	2.91	}^{+	3.86	}$&	0.65	$_{-	0.09	}^{+	0.11	}$&	2.16E-04	&	1.00E+00	&	3190	&	0.29	\\
53508.507	&	7504	&	T	&	266.63	$_{-	7.80	}^{+	7.96	}$&	61.76	$_{-	13.60	}^{+	30.61	}$&	0.76	$_{-	0.09	}^{+	0.19	}$&	2.67E-05	&	3.24E-01	&	10164	&	0.57	\\
\hline
 \multicolumn{10}{c}{XTE J1859+226}\\
 \hline
51485.875	&	6064	&	H	&	269.44	$_{-	6.28	}^{+	5.03	}$&	38.55	$_{-	13.53	}^{+	33.43	}$&	4.91	$_{-	0.71	}^{+	1.57	}$&	2.70E-04	&	7.46E-01	&	824	&	0.21	\\
\hline
 \multicolumn{10}{c}{H 1743-322}\\
 \hline
52803.517	&	6416	&	H	&	240.39	$_{-	4.26	}^{+	3.99	}$&	33.11	$_{-	10.41	}^{+	14.35	}$&	3.01	$_{-	0.36	}^{+	0.59	}$&	1.66E-05	&	2.66E-01	&	1366	&	0.34	\\
52805.425	&	5232	&	H	&	232.38	$_{-	1.38	}^{+	1.41	}$&	7.65	$_{-	3.20	}^{+	6.43	}$&	2.09	$_{-	0.31	}^{+	0.49	}$&	3.13E-04	&	1.00E+00	&	1314	&	0.35	\\
53207.602	&	2608	&	H	&	178.57	$_{-	0.93	}^{+	1.23	}$&	6.01	$_{-	3.80	}^{+	4.09	}$&	6.84	$_{-	1.10	}^{+	1.26	}$&	9.04E-04	&	1.00E+00	&	316	&	0.14	\\
55422.024	&	1440	&	T	&	208.36	$_{-	1.78	}^{+	1.50	}$&	8.04	$_{-	3.38	}^{+	4.68	}$&	5.39	$_{-	0.75	}^{+	0.51	}$&	1.72E-04	&	1.00E+00	&	238	&	0.81	\\
 \hline
 \multicolumn{10}{c}{GX 339-4}\\
 \hline
 52387.556	&	11200	&	H	&	127.70	$_{-	5.57	}^{+	4.16	}$&	44.21	$_{-	10.53	}^{+	26.18	}$&	3.19	$_{-	0.35	}^{+	1.05	}$&	2.95E-06	&	8.12E-02	&	977	&	0.82	\\
52490.229	&	816	&	H	&	278.67	$_{-	3.09	}^{+	3.47	}$&	15.48	$_{-	6.55	}^{+	10.47	}$&	6.32	$_{-	1.01	}^{+	1.45	}$&	8.74E-04	&	1.00E+00	&	1186	&	0.16	\\
52690.841	&	3408	&	T	&	105.35	$_{-	1.50	}^{+	1.77	}$&	8.87	$_{-	3.78	}^{+	4.46	}$&	6.32	$_{-	1.01	}^{+	1.12	}$&	8.45E-04	&	1.00E+00	&	74	&	0.09	\\
54291.443	&	1264	&	T	&	208.44	$_{-	2.68	}^{+	2.81	}$&	14.68	$_{-	5.19	}^{+	8.67	}$&	12.53	$_{-	1.93	}^{+	2.43	}$&	5.98E-04	&	1.00E+00	&	41	&	0.87	\\
 \hline
 \multicolumn{10}{c}{XTE J1752-223}\\
 \hline
 55135.406	&	2896	&	H	&	442.91	$_{-	3.90	}^{+	3.42	}$&	20.23	$_{-	8.27	}^{+	16.19	}$&	7.06	$_{-	1.06	}^{+	1.51	}$&	4.50E-04	&	9.87E-01	&	391	&	0.94	\\
55140.573	&	3424	&	H	&	114.87	$_{-	0.58	}^{+	0.76	}$&	4.50	$_{-	2.39	}^{+	3.42	}$&	2.22	$_{-	0.35	}^{+	0.49	}$&	8.16E-04	&	1.00E+00	&	393	&	0.93	\\
 \hline
 \multicolumn{10}{c}{4U 1630-47}\\
 \hline
 50893.719	&	9936	&	T	&	287.73	$_{-	3.95	}^{+	3.18	}$&	20.17	$_{-	8.60	}^{+	17.73	}$&	2.85	$_{-	0.46	}^{+	0.78	}$&	9.04E-04	&	1.00E+00	&	250	&	0.69	\\
50972.126	&	880	&	T	&	267.55	$_{-	6.59	}^{+	5.32	}$&	32.97	$_{-	12.50	}^{+	14.55	}$&	15.30	$_{-	2.46	}^{+	2.64	}$&	9.35E-04	&	1.00E+00	&	41	&	1.60	\\
52929.482	&	1712	&	H	&	446.72	$_{-	2.95	}^{+	2.39	}$&	14.22	$_{-	7.29	}^{+	8.77	}$&	3.65	$_{-	0.57	}^{+	0.59	}$&	6.41E-04	&	1.00E+00	&	1438	&	0.60	\\
53215.429	&	1488	&	T	&	150.46	$_{-	1.14	}^{+	0.88	}$&	4.91	$_{-	3.09	}^{+	3.36	}$&	1.96	$_{-	0.30	}^{+	0.43	}$&	4.83E-04	&	1.00E+00	&	799	&	0.48	\\

\hline
\end{tabular}
\label{tab:results}
\end{minipage}
\end{table*}

\section{Results}

In this section we examine the results for each source for which HFQPOs have been detected. 
We study in detail
only detections at more than $3\sigma$ significance after keeping number of trials into account.

\subsection{XTE J1550-564}

We selected sixteen features with a single-trial chance probability lower than that corresponding to 3$\sigma$: 1.35$\times 10^{-3}$ (see Tab. \ref{tab:results}). When we took into account the number of trials and considered detections of significance higher than 99\%, we were left with 7 cases. Notice that some detections correspond to a very low count rate, but we nevertheless included them in the sample for completeness. Most of them are at a frequency close to either 180 Hz or 280 Hz, with the exception of the ones with the highest chance probability.
In Figure \ref{fig:hid1550}, we show the HID of all observations of XTE J1550-564 with the detection of HFQPOs (depending on the chance probability). It is clear that all detections correspond to intermediate hardness and it is interesting to note that the observations lie on a diagonal line: the brighter the observed rate, the harder the energy spectrum, with the two detections at a lower frequency corresponding to softer spectra. \revised{The same applies to the gray points, with one exception.}
An examination of the full PDS for these observations indicates that the first one in Tab. \ref{tab:results} corresponds to the brightest observation, where there is a low-frequency QPO (LFQPO) which can be classified as type-A (although Remillard et al 2002a did not classified it at the time). The second observation shows a very clear type-B QPO, while all the others have a type-A QPO. Overall, all observations can be considered part of the soft-intermediate state (SIMS, see Belloni 2010).
The centroid frequencies are clustered around two values: $\sim$180 Hz and $\sim$280 Hz, while their $Q$ value is around 2 for the low-frequency peak and 6-10 for the high-frequency one. The fractional rms of the low-frequency peak is low, around 1\% (total band), as opposed to that of the high-frequency peak which ranges from 3.6\% to 6.7\% (hard band).
Notice that the two last observations belong to a different outburst than the first five, but the QPO centroid frequency is remarkably similar to the others.

\begin{figure}
\begin{center}
\includegraphics[width=8.5cm]{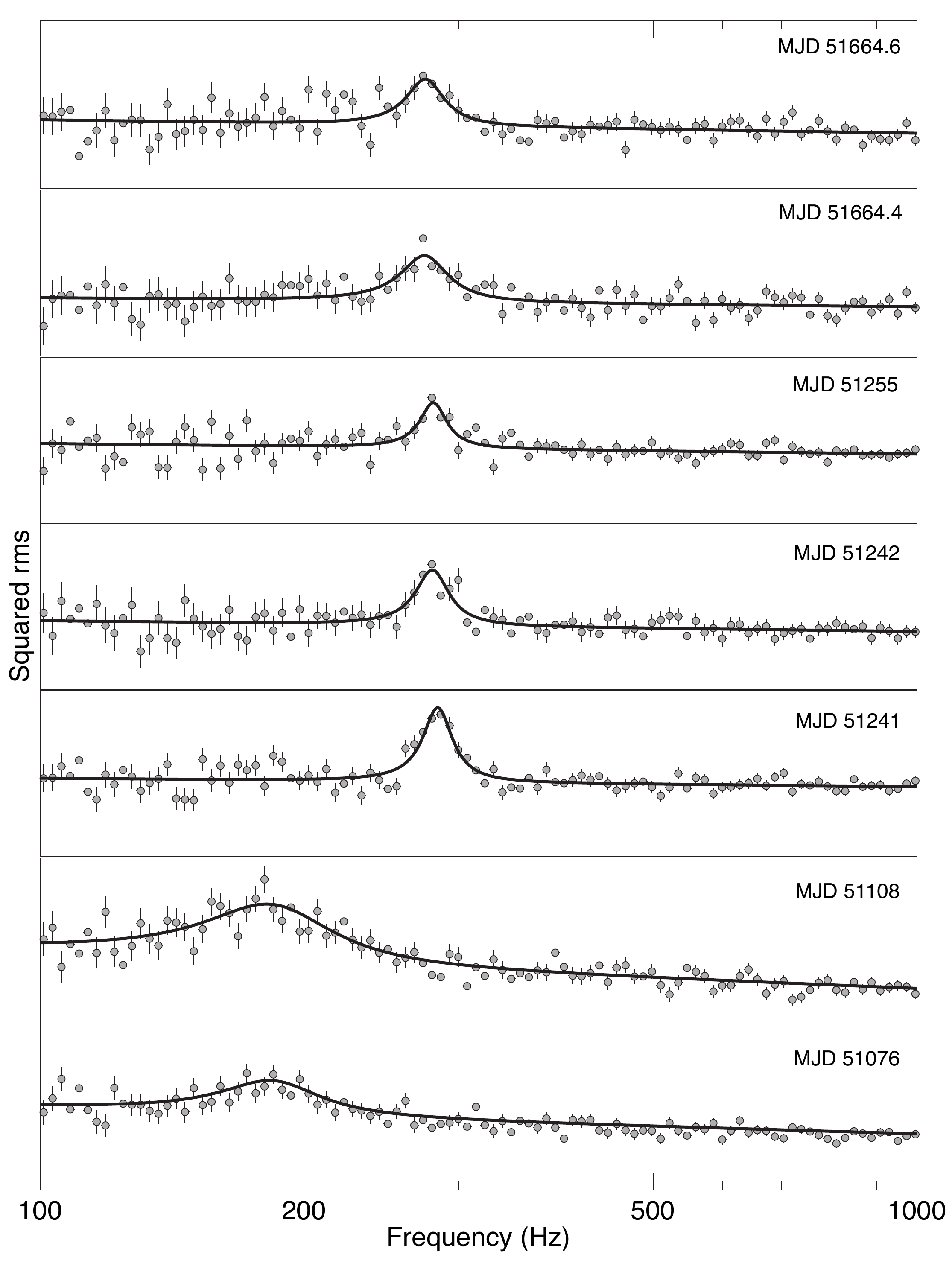}
\end{center}
\caption{PDS of the seven good detections of HFQPO for XTE J1550-564 with the corresponding best fit models.
}
\label{fig:pds1550}
\end{figure}

There are 20 detections reported in Remillard et al. (2002a) from the 1998 outburst of XTE J1550-564. Ten of them appear also in our sample, whereas ten were not detected by our procedure. We checked them all visually and found that many of them resulted in very broad weak excesses superimposed to the tail of a noise component, which led to non significant detections \revised{in our analysis}. Remillard et al. (2002a) do not report any parameter other than the centroid frequency and Q values (from 1 to 15 from their Fig. 7), and it is therefore difficult to make a deeper comparison. 
Remillard et al. (2002a) claim detections at $4\sigma$ level, but they give no indication as to how they estimated the significance.

Miller et al. (2001) report six detections of HFQPO from the 2000 outburst of XTE J1550-564. After taking into account a one-day shift in their Tab. 1, the first four of them correspond to our last four detections. The last two, at slightly lower frequency, are not in our final sample. An inspection of the PDS from those observations show that there is a feature, which our program detected at a significance slightly lower than 3$\sigma$.

\begin{figure}
\begin{center}
\includegraphics[width=8.5cm]{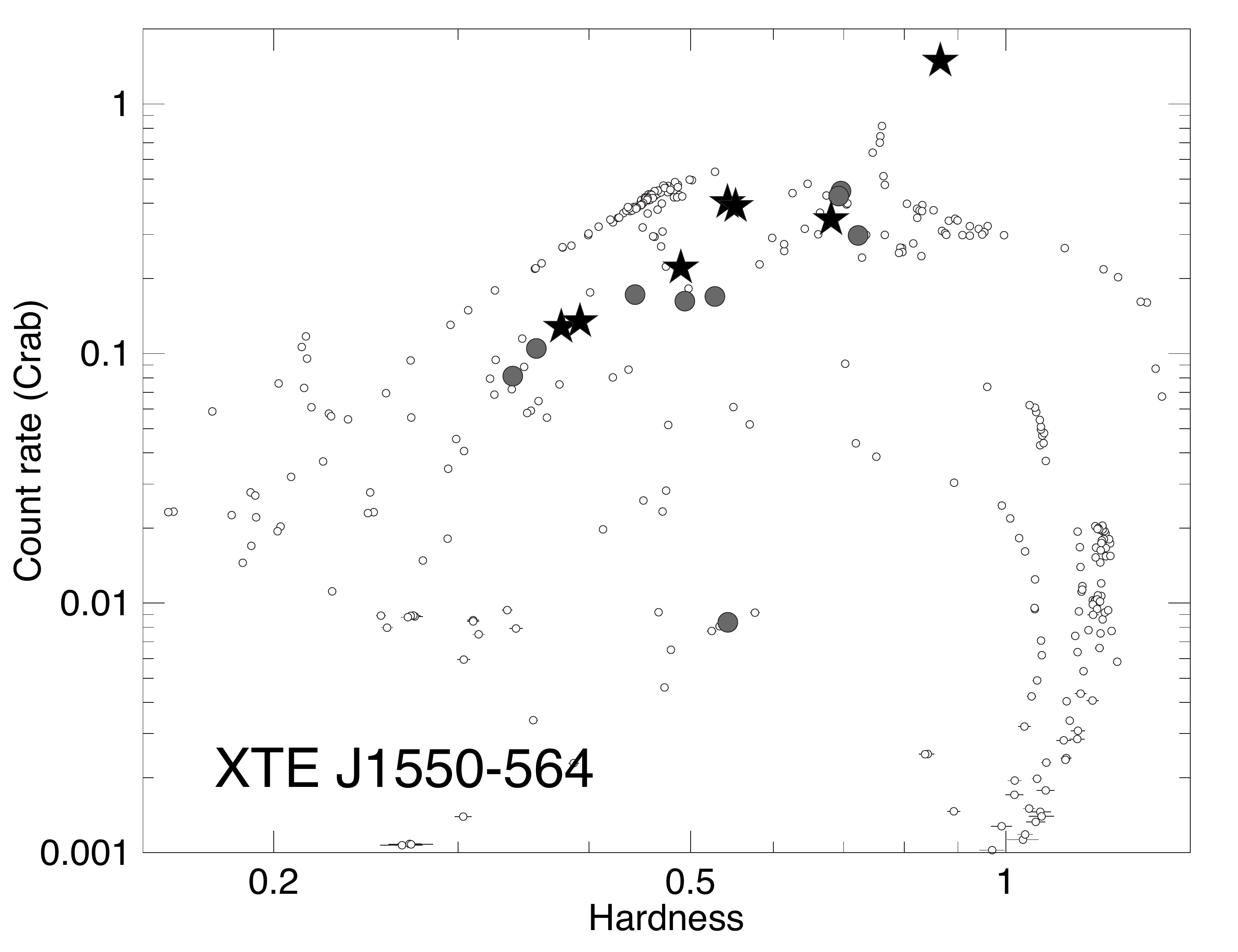}
\end{center}
\caption{Hardness-Intensity Diagram for XTE J1550-564. Each point corresponds to a PCA observation. Gray circles correspond to HFQPO detections with final chance probability larger than 1\%, black stars to those with smaller chance probability.
}
\label{fig:hid1550}
\end{figure}

\subsection{GRO J1655-40}

This is the only other source of our sample with multiple overall detections: 4 out of 11 single-trial detections. The frequency of the QPO is also bimodal: between 273 Hz and 313 Hz for the low-frequency peak ($Q$ between 3 and 4, rms slightly above 1\%) and 446 Hz for the high-frequency peak ($Q$=12 and 6\% rms).
From Fig. \ref{fig:hid1655} it is evident that all detections correspond to the ``anomalous state'' at very high fluxes (see Belloni 2010). The PDS is also typical of that state. As in the case of XTE J1550-564, the HID points are distributed along a diagonal line.
Overall, the four detections with high significance after correction for number of trials show quite a range of frequencies: 274, 290, 313 and 446 Hz.

\begin{figure}
\begin{center}
\includegraphics[width=8.5cm]{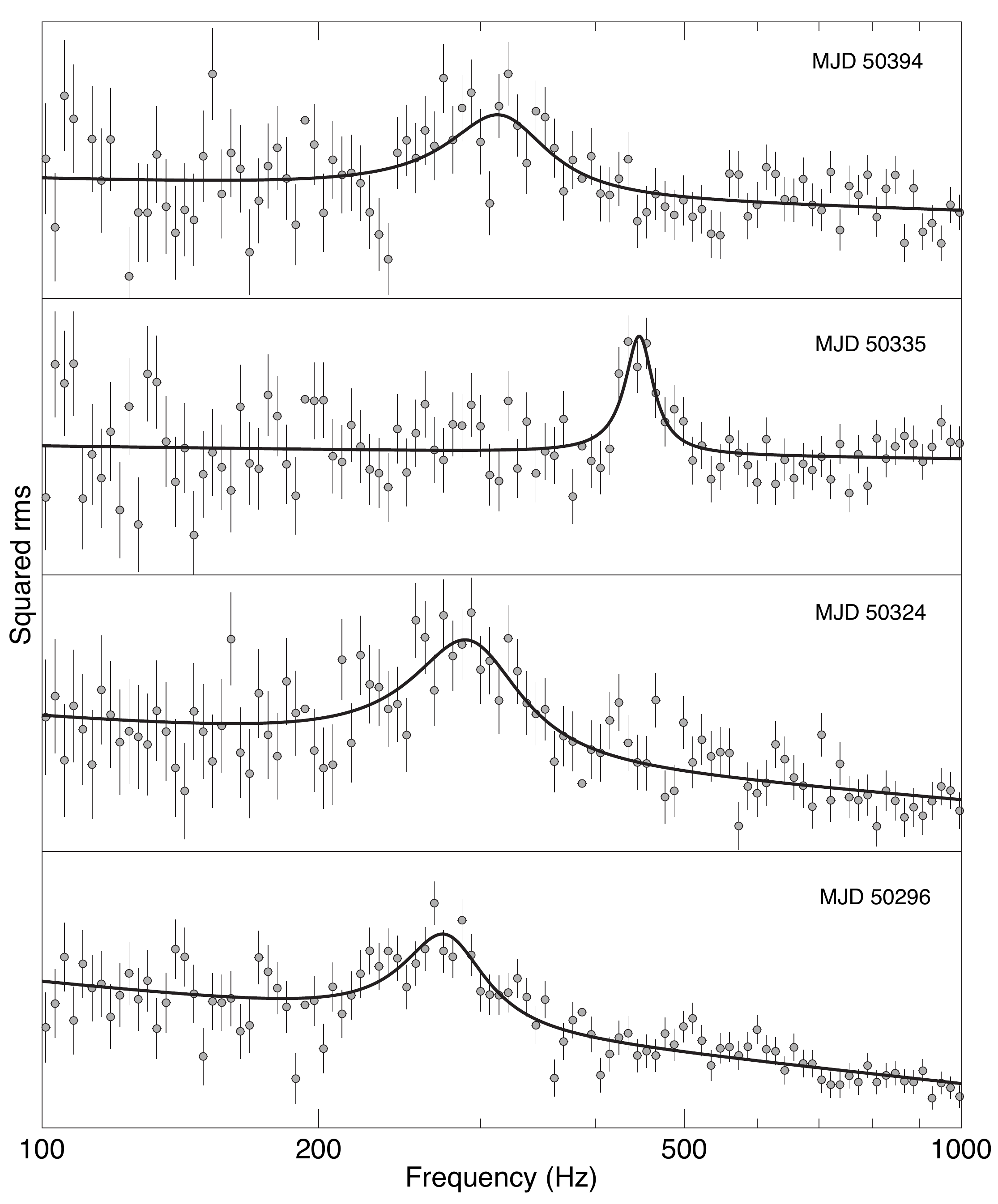}
\end{center}
\caption{PDS of the four good detections of HFQPO for GRO J1655-40 with the corresponding best fit models.
}
\label{fig:pds1550}
\end{figure}

Detections of HFQPO in the 1996 outburst of GRO J1655-40 were reported by Remillard et al. (1999). They detect six features in single observations, in addition to the analysis of sets of combined observations, obviously not comparable with our results. Four of their detections correspond to entries in our Tab. \ref{tab:results}, one corresponds to our 443 Hz observation at MJD 50317, but at \revised{a different frequency (294 Hz); our detection is from the hard band, while theirs is from the total band. In the total band we do detect their signal, but just below 3$\sigma$ single trial. The same is for the observation at MJD 50383.}
As in the case of XTE J1550-564, Remillard et al. (1999) give no details on the detections, nor on the procedure followed.
Two of our detections do not appear in their sample. The observation from MJD 50394 does not appear in their list, while that from MJD 50335 is a very significant peak at 446 Hz in our data, but in the hard band. \revised{Nothing can be seen in the total band, so our result is consistent with theirs. This last one is} the observation where the $\sim$450 Hz QPO was discovered (Strohmayer 2001b). All other observations from Strohmayer (2001b) appear in our sample, although for MJD 50311 \revised{the high-frequency peak found by Strohmayer appears slightly below our 3$\sigma$ level.}

\begin{figure}
\begin{center}
\includegraphics[width=8.5cm]{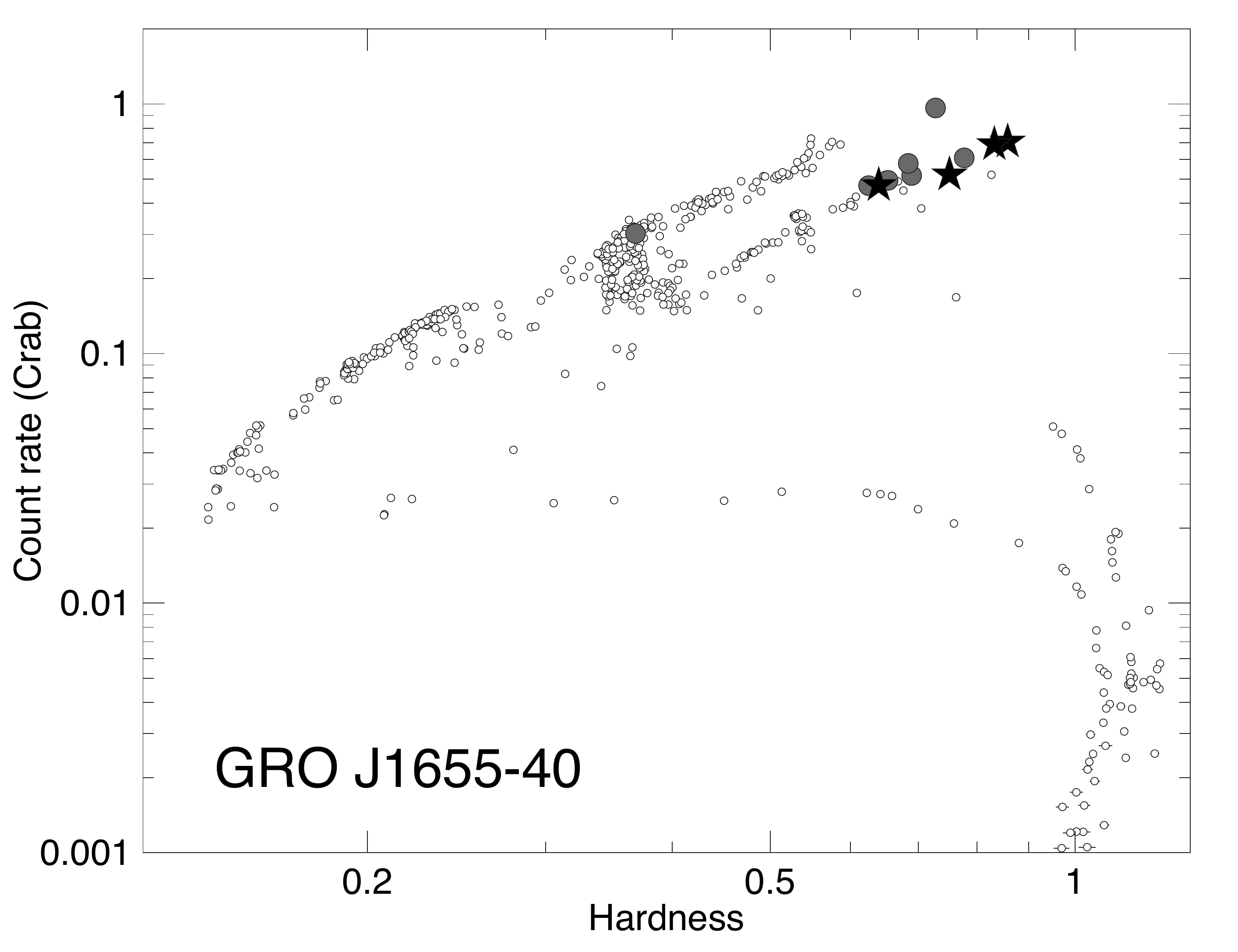}
\end{center}
\caption{Hardness-Intensity Diagram for GRO J1655-40 (see Fig. \ref{fig:hid1550}).
}
\label{fig:hid1655}
\end{figure}

\subsection{XTE J1859+226}

\begin{figure*}
\begin{center}
  \begin{tabular}{@{}ccc@{}}
\includegraphics[width=5.5cm]{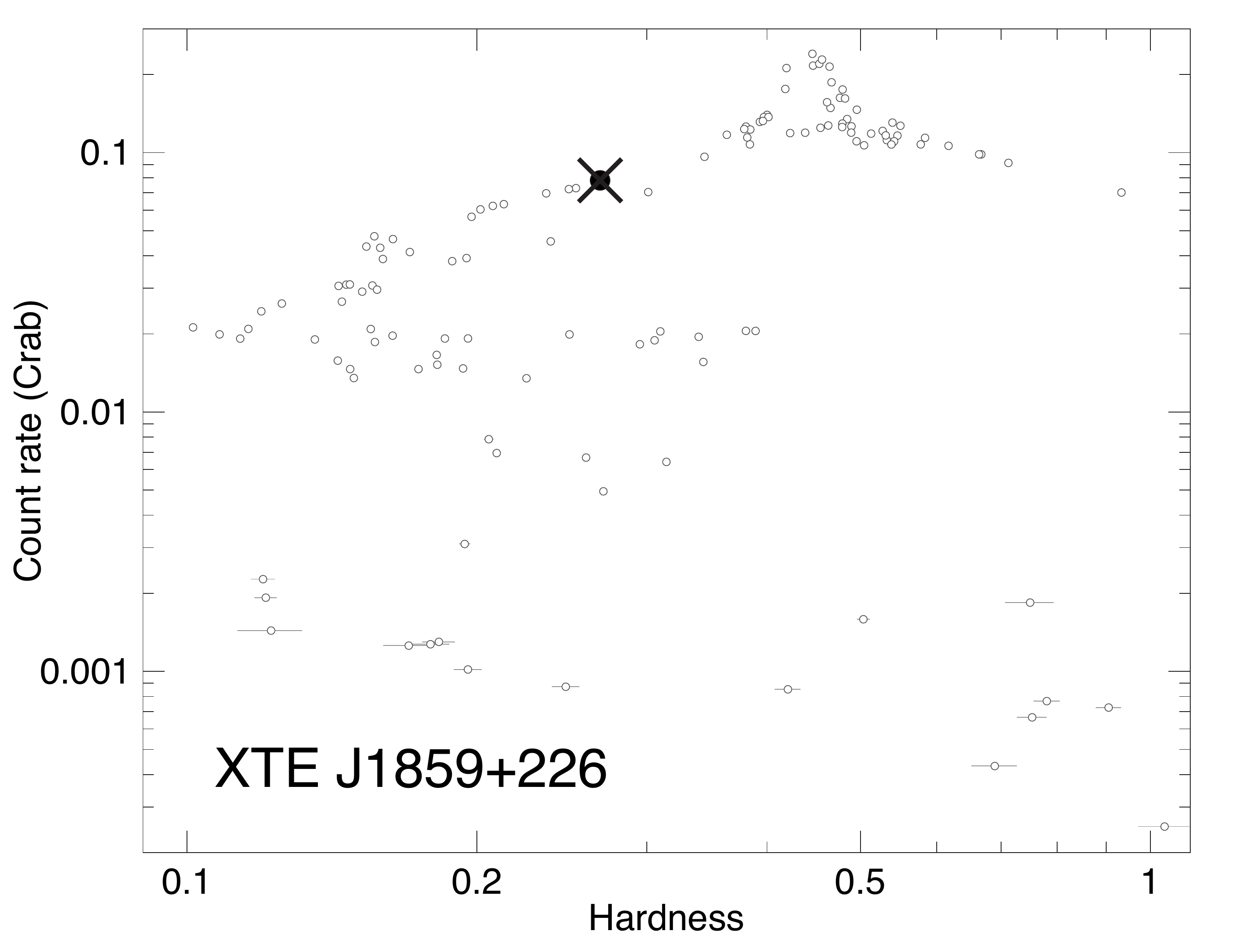} & \includegraphics[width=5.5cm]{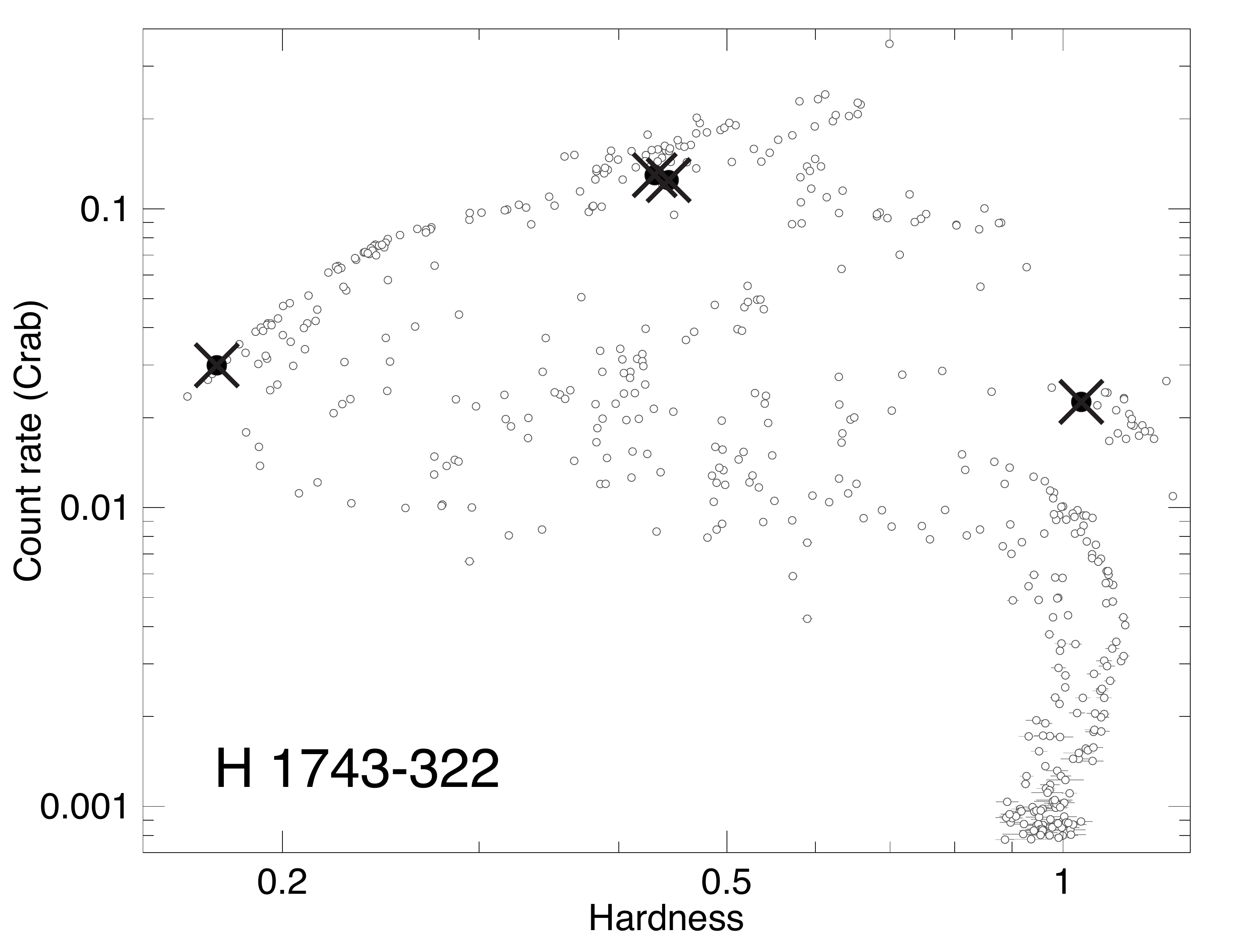} & \includegraphics[width=5.5cm]{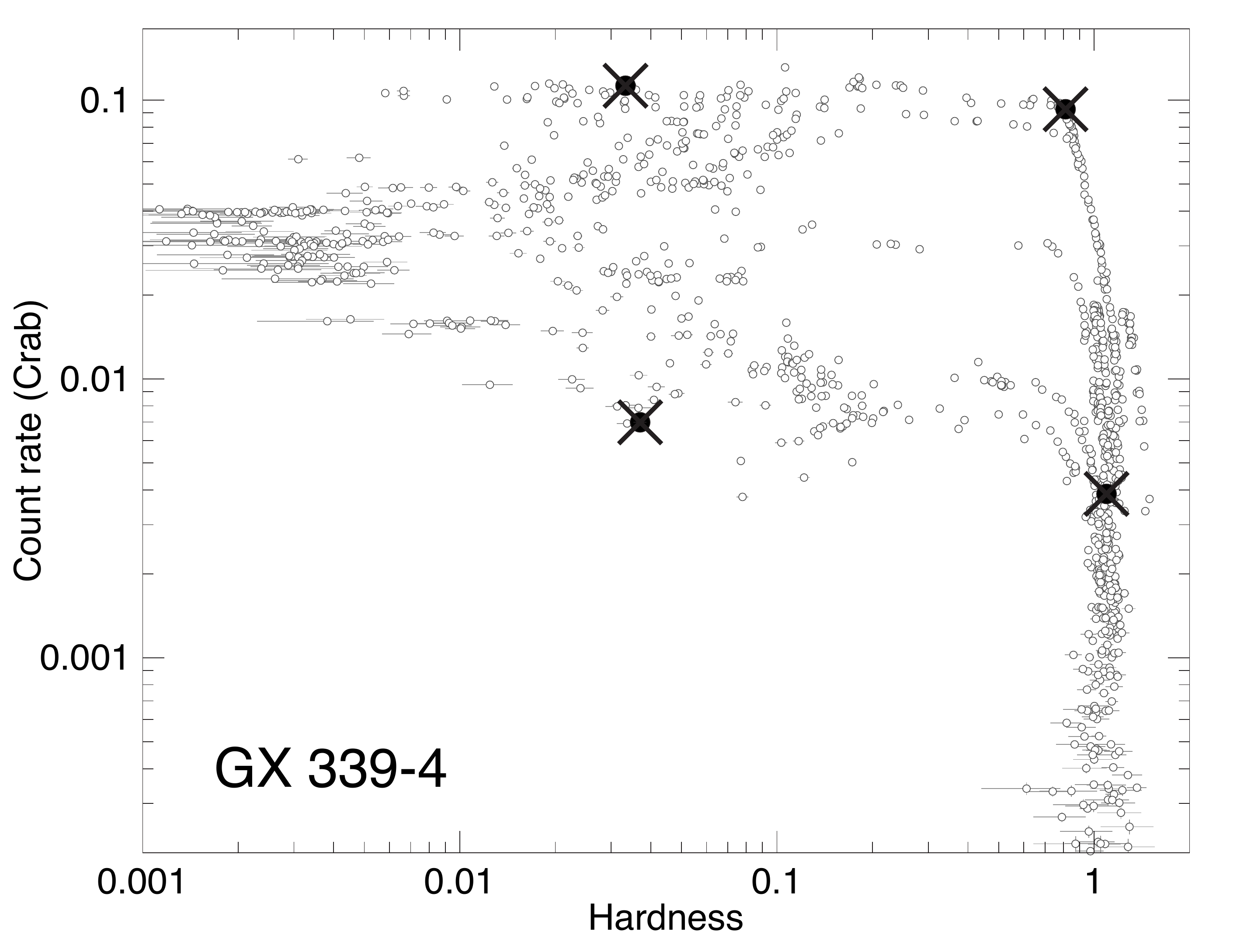}\\
\end{tabular}
  \begin{tabular}{@{}cc@{}}
\includegraphics[width=5.5cm]{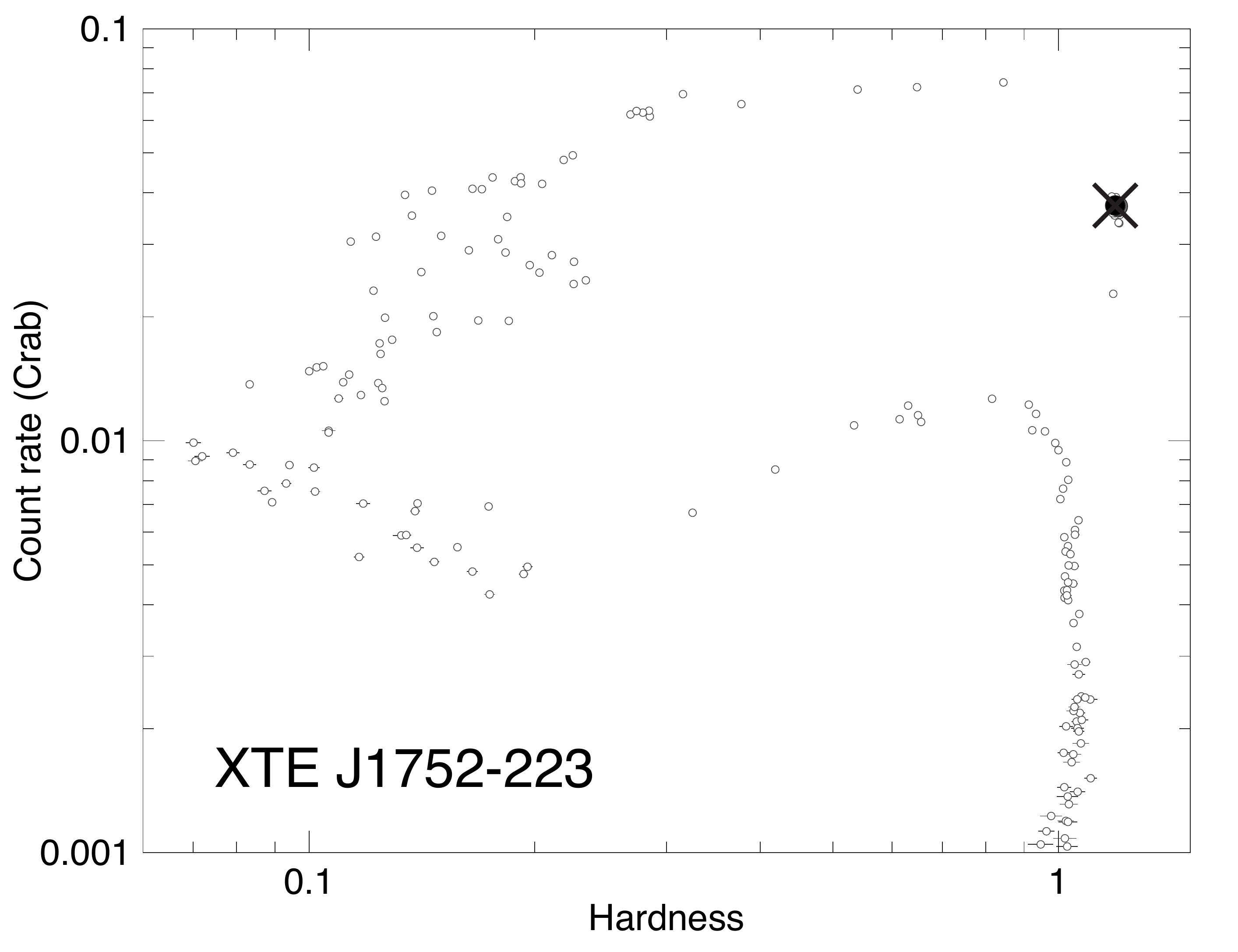} & \includegraphics[width=5.5cm]{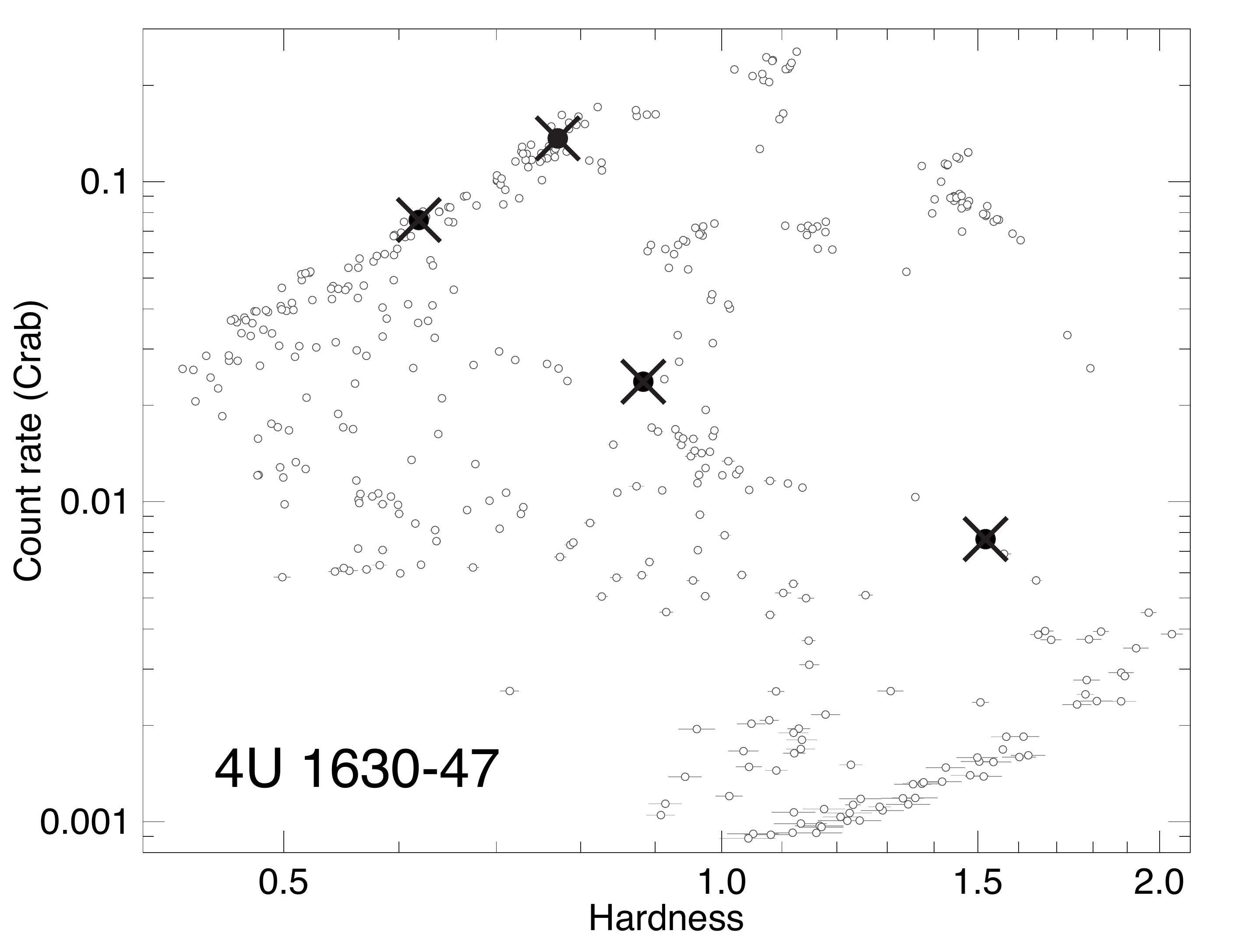} \\
\end{tabular}

\end{center}
\caption{Hardness-Intensity Diagram for the remaining sources for which no HFQPO with chance probability $<1\%$ was found after accounting for the number of trials. Observations with HFQPOs with chance probability $>1\%$ are marked with thick circles and crosses.
}
\label{fig:others}
\end{figure*}

Only one feature appears in our sample, at a single trial probability of 2.7$\times 10^{-4}$, which translates in a very high final chance probability. This feature is very unlikely to be real. \rerevised{The source appears at intermediate hardness in the HID (see Fig. \ref{fig:others}); the presence of a type-B QPO indicates that the source was in the SIMS.}
Cui et al. (2000) report three detections of HFQPOs. None of these QPO appears in our sample. Two of them appear \revised{to be below our $>3\sigma$ single-trial threshold} in their table (the significance is computed as the ratio between normalization and its lower-side error, while using fractional rms as done by Cui et al. (2000) increases the value by a factor of two because of the error propagation), while the third one is \revised{a very broad feature which is automatically discarded by our procedure.}

\subsection{H 1743-322}

\rerevised{We detect four QPOs, with high final chance probability.  
From Fig. \ref{fig:others} we can see that in one case the source was in a very soft HSS, in one case the source was rather hard (the full PDS shows a type-C QPO, from which we conclude the source was in the HIMS), and in two cases the source showed intermediate hardness, and was in a similar location in the HID as the ones discussed for GRO J1655-40 and XTE J1550-564. In these last two cases the full PDS shows a strong type-B QPO, and therefore the source was in the SIMS.
}
HFQPOs were reported from H 1743-322 by Homan et al. (2005) at 163 and 240 Hz from the observations of MJD 52787. They do not appear in our sample, but accumulation of more than one observation is needed to obtain a detection, so our results are not inconsistent with theirs. The same applies to the detection of Remillard et al. (2006).

\subsection{GX 339-4 and XTE J1752-223}

For GX 339-4, the fur detections form a square in the HID (see Fig. \ref{fig:others}). The two hard ones are obviously LHS. The two soft ones are HSS. For XTE J1752-223 the two detections are at the same high value of hardness, in the LHS.
\rerevised{In these two case the chance probability was, respectively, 0.23 and 0.035. These are most likely statistical fluctuations and therefore we do not discuss them further.}

\subsection{4U 1630-47}

None of the four peaks in Tab. \ref{tab:results} appear to be statistically significant after the (large) number of trials is taken into account. 
\rerevised{Two peaks are detected when the source flux was low: in one case the source was in the LHS, while in the other case the full PDS shows very little noise, making it difficult to classify the state. The remaining two observations lie along a diagonal branch in the HID and also show low noise level.
}
High-frequency features in this source were reported by Klein-Wolt et al. (2004). Those are very broad features which depend crucially on the subtraction of Poissonian noise. However, all of them were obtained after averaging observations corresponding to different intervals in the 1998 outburst and therefore cannot appear in our sample.

\subsection{XTE J1650-500}

We did not obtain detections for this source. Homan et al. (2003) report high-frequency features, some of which very broad like in the case of 4U 1630-47. Once again, considerable averaging of more observations has been performed, which explains the absence of detections in our sample.

\section{Discussion}

We analyzed 7108 RXTE/PCA observations of 22 black-hole transients searching for HFQPO. From the whole sample, 
after taking into account the number of trials due to the many observations for each source and the number of independent frequencies sampled,
we obtain only 11 detections with a chance probability less than 1\%. These detections belong to two sources : XTE J1550-564 (7) and GRO J1655-40 (4). The small number of detections (including single-trial detections at significance higher than 3$\sigma$ there are only 42), confirms that these are very elusive signals, as already indicated by the \revised{small number of} detections reported in the literature. A few additional detections were obtained by other authors by averaging observations according to hardness, but those cases will not be discussed here. From our detections, a number of aspects can be discussed.

All detections correspond to PDS from very specific states. Six out of seven detections for XTE J1550-564 are from the SIMS, while the last one corresponds to the high-soft state (HSS). One observation has a clear type-B QPO, three have a type-A QPO and one corresponds to the bright flare in 1998, where a LFQPO can be tentatively identified with type-A (see Remillard et al. 2002a). No type-C QPO is detected simultaneous to one of the HFQPO. The first five HFQPO are from the 1998 outburst, the last two from the 2000 outburst (see Rodr\'\i guez et al. 2004).
The four detections from GRO J1655-40 correspond to the ``anomalous state'' described in Belloni (2010). Again, no type-C QPO can be identified in those four PDS. These detections come all from the 1996 outburst (Remillard et al. 1999).
From the HIDs in Figs. \ref{fig:hid1550} and \ref{fig:hid1655}, we see that the points corresponding to the HFQPO detections in both sources are distributed diagonally. However, it is difficult to relate this to precise spectral characteristics.

For XTE J1550-564, the centroid frequencies of the detected QPOs appear to cluster around two values: $\sim 180$ Hz and $\sim$280 Hz. Interestingly, the low-frequency detections come from the total energy band, while the high-frequency detections come from the hard band, indicating that the high-frequency oscillation has a harder spectrum. Moreover, inspection of Fig. \ref{fig:pds1550} shows that the low-frequency peaks are clearly broader than the high-frequency ones.

For GRO J1655-401, the dichotomy 300-450 Hz seems to be less sharp. Strohmayer (2001) detected the $\sim$450 Hz feature simultaneous with the 300 Hz one when averaging observations, showing that the two peaks correspond to different time scales. We see also frequencies ($\sim$270 and $\sim$290 Hz) significantly lower than 300 Hz. Also in this source the high-frequency detection is significantly narrower than those at lower frequencies.

The fact that the few available detections appear at more or less the same frequencies for XTE J1550-565, even for different outbursts, and that the upper peak is narrower than the lower make a comparison with kHz QPO in neutron-star binaries difficult. There, the centroid frequencies are observed to vary in a way consistent with a random walk (see Belloni et al. 2005, 2007) and the upper peak is broader than the lower one (see e.g. Barret et al. 2005).

Two sources which are known to show HFQPOs were excluded from the analysis. GRS 1915+105 is a very peculiar system with many observations and many HFQPO detections. We will present our results on this source in a forthcoming paper. 
IGR J17091-3624 is a new ``twin'' of GRS 1915+105, in the sense that it is the only other known source to display its unusual variability. Results were already published in Altamirano et al. (2012).

It is interesting to compare our results with those of kHz QPOs in neutron-star LMXBs (see van der Klis 2006 for a review). The neutron-star system 4U 1636-53 was observed regularly during the last few years of the RXTE lifetime, yielding a solid database from which to extract statistical information. In this source, $\sim$40\% of the observations led to the detection of at least one high-frequency feature (see Sanna et al. 2012) as opposed to 0.15\% here. 

Revnivtsev \& Sunyaev (2000) showed that there is a systematic difference in power level between black-hole and neutron-star sources in the hard state, attributing it to the presence of a radiation-dominated spreading layer in the latter, the so-called boundary layer. As already suggested by these authors, the presence/absence of strong high-frequency QPOs can be part of the same picture. Following this line, Gilfanov et al. (2003) and Gilfanov \& Revnivtsev (2005) analysed low-frequency and kHz QPOs in neutron-star binaries and concluded that the oscillations are caused by variations in the luminosity of the boundary layer. In this framework, it is the diminishing relative importance of the boundary-layer emission which weakens kHz QPO as they move to very high frequencies (see M\' endez 2006 for additional discussion).
Moreover, our HFQPOs are detected in intermediate states. Belloni et al. (2007) and Sanna et al. (2012) showed that in the neutron-star LMXB 4U 1636-53 kHz QPOs are not found in the observations when the source is the hardest, at intermediate hardness mostly the upper kHz QPO is detected and in the softest observations mostly the lower kHz QPO is present. It is difficult to make a comparison with our case: that no kHz QPOs are detected in the hard state is consistent with the black-hole case, while a detailed comparison between soft states remains to be done.

\begin{figure}
\begin{center}
\includegraphics[width=8.5cm]{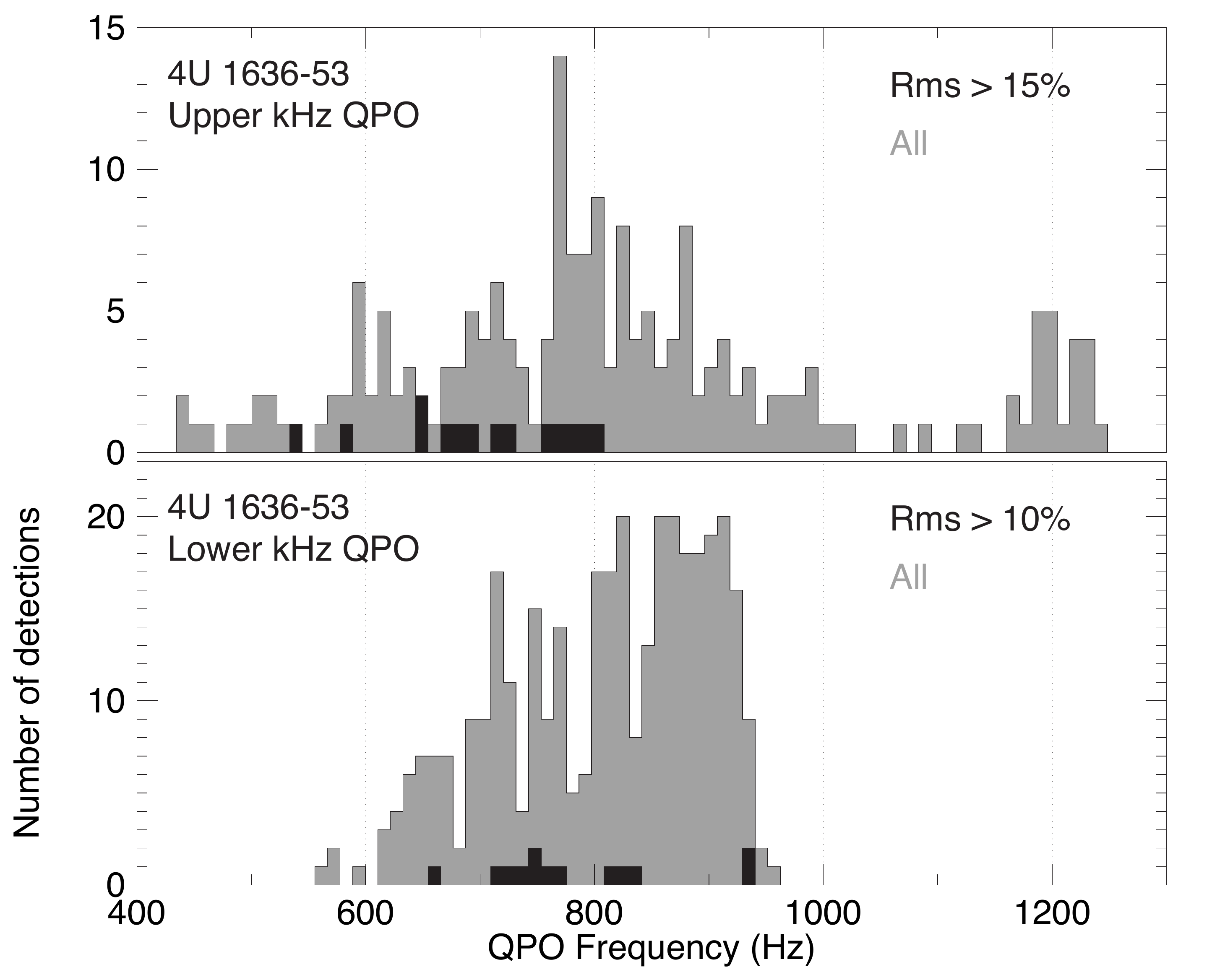}
\end{center}
\caption{Frequency distributions for all detections of upper (top panel) and lower (bottom panel) kHz QPOs in the NS LMXB 4U 1636-53 (Sanna et al. 2012) (gray histograms). The black histograms include only detections with rms $>$15\% for the upper and $>$10\% for the lower.
}
\label{fig:4u1636}
\end{figure}

The detection of only a few specific frequencies in the case of black-hole HFQPOs is something that does not apply to kHz QPOs in neutron-star binaries. The frequencies of kHz QPO are distributed along a very broad range (see van der Klis 2006, M\'endez 2006, Belloni et al. 2007, Barret et al. 2005, 2007). 
\revised{
KHz QPOs in neutron stars are usually strong, more easily detected than HFQPO, which then leads
to a much larger number of detections. Moreover, at most frequencies the lower kHz peak is narrower and more significant than the upper one. 
The question is whether the frequencies of the oscillations can be understood in terms of the same physical mechanisms in the two classes of sources.
In order to check what frequencies would be observed if this was the case, in Fig. \ref{fig:4u1636} we plot the distribution of upper and lower kHz QPOs in the full RXTE database of 4U 1636-53 (data from Sanna et al. 2012). In both panels, the black histogram contains only the strongest detected peaks (we chose an rms threshold of 15\% and 10\% for the upper and lower peaks respectively in order to select only a handful of detections). A comparison with the HFQPOs of XTE J1550-564 is interesting. There, the lower-frequency HFQPO is weaker and broader than the higher-frequency one. Unfortunately, the only reported case of a double-detection in XTE J1550-564, obtained through averaging of many observations, does not allow a good determination of the quality factor of the two peaks (Miller et al. 2005).
The black histograms show that on average the strongest upper kHz QPOs in 4U 1636-53 appear at a lower frequency than the strongest lower kHz QPOs. This is consistent with the spread of the frequencies we find in XTE J1550-564.
Clearly, the limited number of detections of XTE J1550-564 does not allow a more precise comparison.
}

The width of the QPO peaks is another interesting quantity of a comparison with neutron-star binaries. In our detections,  the $Q$ factor of the upper-frequency peak is higher than that of the lower-frequency one. In neutron-star binaries it is generally the opposite (see e.g. Barret et al. 2007). This is a clear difference between the neutron-star and black-hole cases. If the process originating the oscillations is the same of the two classes of courses, the presence of the boundary layer is likely to be the reason for this difference, yielding another observational constraint for theoretical modeling (see M\' endez 2006).

Overall, we detected eleven significant features after analyzing around 7000 observations, resulting in a detection efficiency of $\sim10^{-3}$. Of course, since the HFQPOs are detected only in intermediate states and these states are the short-lived ones, observing with new instruments and concentrating on the most likely periods during outbursts can increase the probability of detection. In other words, to detect HFQPOs one does not need a good coverage of a full outburst, but only of the intermediate states. The LAXPC instrument on board the upcoming ASTROSAT mission (Agrawal 1996) will increase the effective area of the PCA above 10 keV, where the HFQPOs are stronger (see Morgan et al. 2001), but in order to be effective ASTROSAT will have to concentrate observations during bright intermediate states. The LOFT mission (Feroci et al. 2010), under evaluation by ESA at the time of writing, is planned with a very large effective area (around 12 m$^2$) and will allow sampling of these weak signals in all states.

\section*{Acknowledgements}

The research leading to these results has received funding from the European CommunityÕs Seventh Framework Programme (FP7/2007-2013) under grant agreement number ITN 215212 ÔBlack Hole UniverseÕ. 
\revised{This paper was written during an extended stay of TMB at The University of Southampton, funded by a Leverhulme Trust Visiting Professorship.} TMB thanks S. Motta for her help with source states.

\end{document}